\documentstyle[prd,aps,rotating]{revtex}
\input epsf
\begin{document}
\onecolumn
%\vfill
\begin{titlepage}
{\flushright      WATPHYS-TH99/01}\\
{\flushright ITP NSF-ITP-99-21}\\
\begin{center}
{\Large \bf Misner String Entropy} \\ \vspace{2cm}
R.B. Mann\footnotemark\footnotetext{email: 
rbmann@itp.ucsb.edu \\
on leave from Dept. of Physics, University of Waterloo, Waterloo
Ont. Canada N2L 3G1} \\
\vspace{1cm}
Institute for Theoretical Physics\\
Dept. of Physics\\
University of California Santa Barbara\\
Santa Barbara, CA USA 93106\\
\vspace{2cm}
\vspace{2cm}
\today\\
\end{center}
\begin{abstract}
I show that gravitational entropy can be ascribed to
spacetimes containing Misner strings (the gravitational analogues of Dirac strings),
even in the absence of any other event horizon (or bolt) structures. This result follows
from an extension of proposals for evaluating the stress-energy of a gravitational system
which are motivated by the AdS/CFT correspondence.
\end{abstract}

\end{titlepage}
\onecolumn

The idea that one can associate entropy with certain  gravitational field configurations
has a long history, dating back to Beckenstein's suggestion \cite{beck} that the area of the event
horizon of a black hole is proportional to its physical entropy and to Hawking's
demonstration \cite{hawk1} that black holes can radiate with a black body spectrum at non-zero
temperature once quantum effects are taken into account. The relationship between
the entropy of a black hole and the area of its event horizon is quite robust,
being applicable in any number of dimensions and extendable to black branes and dilatonic 
variants \cite{qgag}.

Hawking and Hunter have recently suggested \cite{Hawkhunt} that gravitational entropy arises whenever it is
not possible to foliate a given spacetime in the Euclidean regime by a family of surfaces 
of constant time $\tau$. Such breakdowns in foliation can occur if the topology of the Euclidean 
spacetime is not trivial -- in particular when the U(1) isometry group 
associated with the (Euclidean) timelike Killing vector $\xi=\partial/\partial\tau$
has a fixed point set of even co-dimension.  If the fixed point set has co-dimension $d-2$
then the usual relationship between entropy and area holds. However if there exist
additional fixed point sets with lower co-dimensionality then the relationship between
entropy and area  is generalized. 
This has been explictly demonstrated in Taub-bolt
and Taub-bolt-Anti de Sitter (AdS) spacetimes in 4 dimensions.  Both of these spacetimes have 
$2$-dimensional fixed point sets (i.e. event horizons, referred to as ``bolts'' \cite{gibh}) 
as well as $0$-dimensional fixed point sets (referred to as ``nuts'') at which the
orbits of the U(1) isometry group become singular.  In the latter case the singularities
are of dimension 1 in the orbit space (and of dimension 2 in the Euclidean spacetime). They are 
the gravitational analog of Dirac string singularities and are referred to as Misner strings.  
The presence of Misner strings implies that one must include in the entropy budget
not only the areas of the bolts, but also those of the strings and the Hamiltonians
evaluated on the strings. These latter quantities are divergent;
by subtracting their analogous values in spacetimes with appropriately matched asymptotic structure
with no bolts present, finite values for the entropies for each spacetime may
be obtained \cite{Hawkhunt,HPH,Chamb}. 

The purpose of this paper is to carry such arguments a step further by demonstrating that
Misner strings themselves have an intrinsic entropy.  One need not have a bolt present for
a spacetime to have gravitational entropy.  This result follows from recent work on the
conjectured AdS/CFT duality, which equates the bulk
gravitational action of an asymptotically AdS spacetime with the quantum effective action
of a conformal field theory (CFT) defined on the AdS boundary. Inspired by this correspondence,
Balasubramanian and Kraus \cite{vjk} have proposed adding a term to the boundary at infinity (understood
as boundary at some radius $r$ in the large-$r$ limit) which is a functional only of curvature 
invariants of the induced metric on the boundary (see also ref. \cite{hyun}). The addition of such terms does not affect
the bulk equations of motion because they are intrinsic invariants of the boundary metric.
In four dimensions only two invariants exist which (due to dimensionality) can contribute to the 
Hamiltonian at infinity, and their coefficients are uniquely fixed by demanding that 
it be finite for Schwarzchild-AdS spacetime. As I shall demonstrate, this choice of boundary term 
is sufficient to compute finite values for the actions, Hamiltonian, and entropies for both Taub-bolt-AdS
and Taub-NUT-AdS spacetimes. A recent observation by Lau \cite{lau} permits 
an extension of this boundary term to asymptotically flat cases, and I shall 
demonstrate how it may be used to compute entropies of the Taub-bolt and Taub-NUT 
spacetimes. Taub-NUT and Taub-NUT-AdS are the
first examples of spacetimes without  bolts that have gravitational entropy.  

The general relationship between entropy and area follows from the
thermo\-dynamic definition 
\begin{equation}\label{e1}
\log Z = S - \beta H_\infty
\end{equation}
combined with the path-integral derivation that the Euclidean action $I = -\log Z$ to lowest order,
where $Z$ is the partition function of an ensemble
$$
Z = \int[Dg][D\Phi]\exp[-I(g,\Phi)]
$$
with the path integral taken over all metrics $g$ and matter fields $\Phi$ that are 
appropriately indentified under the period $\beta$ of $\tau$. 
The entropy is then the difference between the value 
the action would have ($\beta H_\infty$) if there were no breakdown of foliation and its 
actual value. In order to compute this difference, 
one can remove small neighbourhoods $N_\epsilon^i$ of the fixed point sets and strings,
so that $I= I_{M_\epsilon^i}-\sum_i I_{N_\epsilon^i}$. Rewriting each $I_{M_\epsilon^i}$
in Hamiltonian form, taking care to include the additional surface terms
due to these new boundaries, one finds that the Hamiltonian in the bulk and at the fixed point
set boundaries vanishes; the only non-zero contributions are from the boundary at infinity and
from the boundary along the strings, and so $I_{M_\epsilon^i} = \beta \sum(H_\infty + H_{{MS}})$.
In the limit that the $N_\epsilon^i$ vanish, one finds \cite{Hawkhunt,HPH} 
$I_{N_\epsilon^i}
=\frac{1}{4}\sum(\cal{A}_{\mbox{bolt}}+\cal{A}_{{MS}})$; the bulk parts are zero but 
their surface terms are non-vanishing and yield the one-quarter of the areas of the neighbourhoods 
removed, i.e. the bolts and the strings.

Hence
\begin{equation}\label{e2}
S = \beta H_\infty - I = \frac{1}{4}\sum\left(\cal{A}_{\mbox{bolt}}+\cal{A}_{{MS}}\right)
      -\beta\sum\left(H_{{MS}}\right)
\end{equation}
is the generalized entropy formula. The action is generally taken to be a linear combination
of a volume (or bulk) term
\begin{equation}\label{e3}
I_v = -\frac{1}{16\pi}\int_M d^dx\sqrt{g}\left(R+2\Lambda+{\cal L}(\Phi)\right)
\end{equation}
and a boundary term
\begin{equation}\label{e4}
I_b = -\frac{1}{8\pi}\int_{\partial M} d^{d-1}x\sqrt{\gamma}\Theta(\gamma)
\end{equation}
chosen so that the variational principle is well-defined. The Euclidean manifold $M$ has metric $g_{\mu\nu}$,
covariant derivative $\nabla_\mu$, and time coordinate $\tau$ which foliates $M$ into non-singular
hypersurfaces $\Sigma_\tau$ with unit normal $u_{\mu}$. $\Theta$ is the trace of 
the extrinsic curvature $\Theta^{\mu\nu}$ of any boundary(ies) $\partial M$ of
the manifold $M$, with induced metric(s) $\gamma$; the manifold can have  internal boundary components
as well as a boundary at infinity, although only the latter will be needed in what follows.
Here ${\cal L}(\Phi)$ is the matter Lagrangian and $\Lambda$ the cosmological constant.

In general $I_v$ and $I_b$ are both divergent when evaluated on solutions, as are the string area and
Hamiltonian terms in (\ref{e2}) above.  The general method for treating this problem is to
consider the values of these quantities relative to some reference background spacetime whose
boundary(ies) have the same induced metric(s) as those in the original spacetime \cite{BY,BCM,HHor}.  
This background is chosen so that its topological properties are compatible with the solution and the 
solution approaches it
sufficiently rapidly at infinity. The reference spacetime is then interpreted as the vacuum for that sector
of the quantum theory.  Unfortunately this procedure suffers from certain drawbacks.  
It is not always possible 
to embed a boundary with a given induced metric into the reference background.  Furthermore the reference
spacetime is not necessarily unique, nor does it necessarily satisfy the invariance properties one
might require of certain thermodynamic quantities \cite{CCM}.

An attractive resolution to this difficulty in AdS spacetimes involves adding a term $I_{ct}$ to the
action which is a functional only of boundary curvature invariants.  As noted above, such a prescription
is inspired by the conjectured AdS/CFT correspondence: divergences which appear in the stress-energy tensor
of the CFT on the boundary are just the standard ultraviolet divergences of quantum field theory and may
be removed by adding counterterms to the action which depend only on the intrinsic geometry of the boundary,
obviating the need to include any reference spacetime at all.  Quantities such as energy, entropy, etc.
can then be understood as intrinsically defined for a given spacetime, as opposed to being defined relative
to some abstract (and non-unique) background, although this latter option is still available.  In four
dimensions \cite{vjk}
\begin{equation}\label{e5}
I_{ct(AdS)} = \frac{2}{\ell}\frac{1}{8\pi}\int_{\partial M_\infty} 
d^{3}x\sqrt{\gamma}\left(A + B\frac{\ell^2}{4} R(\gamma)\right) 
\end{equation}
where $\ell^2 = 3/|\Lambda|$ and  
$A$ and $B$ are dimensionless coefficients chosen so that the 
total action $I=I_v + I_b + I_{ct(Ads)}$ is finite. Although
other counterterms (of higher mass dimension) may be added to $I_{ct}$, they will make no contribution
to the evaluation of the action or Hamiltonian due to the rate at which they
decrease toward spatial infinity.
Although the prescription (\ref{e5}) applies only to spacetimes which 
are asymptotically AdS, it may be generalized to 
\begin{equation}\label{e7}
I_{ct} = \frac{2A}{\ell}\frac{1}{8\pi}\int_{\partial M_\infty} 
d^{3}x\sqrt{\gamma}\sqrt{1 + B\frac{\ell^2}{2A} R(\gamma)} 
\end{equation}
which is equivalent to (\ref{e5}) for small $\ell$ (as well as fixed $\ell$ and large
mean boundary radius),  and which (formally)
has a well-defined limit for vanishing
cosmological constant ($\ell\to\infty$).  In this limit (\ref{e7}) 
reduces to a prescription given by
Lau \cite{lau}. 

Empowered with the preceding prescription, the entropy of a given spacetime is then
\begin{equation}\label{e8}
S = I_v + I_b + I_{ct}-\beta H_\infty 
\end{equation}
where all quantities are evaluated on a given solution. By taking the variation of the action with
respect to the boundary metric $\gamma_{\mu\nu}$ at infinity, it is straight\-forward to show that
\begin{equation}\label{e9}
H_\infty = \frac{1}{8\pi}\left[\Theta^{\mu\nu}-\Theta \gamma^{\mu\nu} 
+ \frac{2}{\sqrt{\gamma}}\frac{\delta I_{ct}}{\gamma_{\mu\nu}}\right]u_{\mu}\xi_{\nu}
\end{equation}
where $\xi^\mu$ is a timelike killing vector on the boundary at infinity. Note that in general
it is not coincident with the unit normal $u^\mu$ \cite{BCM}. The extrinsic curvature 
\begin{equation}\label{e10}
\Theta^{\mu\nu} = -\frac{1}{2}\left(\nabla^\mu n^\nu + \nabla^\nu n^\mu\right)
\end{equation}
where $n^\mu$ is the unit normal to the boundary at infinity. 

The coefficients $A$ and $B$ may be determined by considering the 
Schwarzchild-AdS (SAdS) solution
\begin{equation}\label{e11}
ds^2 = (\frac{r^2}{\ell^2}+1-\frac{2M}{r})d\tau^2 + \frac{dr^2}{\frac{r^2}{\ell^2}+1-\frac{2M}{r}}
+r^2 d\Omega^2
\end{equation}
which has a bolt at $r=r_+$, where $\frac{r_+^3}{\ell^2}+r_+-2M=0$ and
$d\Omega^2=d\theta^2+\sin^2(\theta)d\phi^2$. Foliation of the spacetime
breaks down at this point, and regularity of the solution implies that the period of $\tau$ is
$\beta_{SAdS} = (3r_+^2 +\ell^2)/(4\pi\ell^2 r_+)$.  Finiteness of the total action implies
that $A=B=1$, yielding
\begin{equation}\label{e12}
I_{SAdS}=\pi r_+^2 \frac{\ell^2-r_+^2}{3 r_+^2+\ell^2}  
\end{equation}
for the total SAdS action.  This result is the same that one would obtain by taking the
difference between the SAdS action without counterterm and the action of pure anti de Sitter
space whose time coordinate has the same periodicity.  Using $A=B=1$ in (\ref{e9}) yields
$H_{SAdS,\infty}=M=(r_+^3+\ell^2 r_+)/2\ell^2$, and so from
(\ref{e8}) the entropy of SAdS spacetime is
\begin{equation}\label{e12a}
S_{SAdS}= \pi r_+^2 \qquad .
\end{equation}

Setting $A=B=1$ in (\ref{e7}), equations (\ref{e8}) and (\ref{e9}) may be used to
evaluate the total action, the Hamiltonian at infinity and the entropy of 
flat space, Schwarzchild,
Taub-NUT, Taub-bolt, and Topological black hole spacetimes, whose metrics are of the form
\begin{equation}\label{e13}
ds^2= V(r)(d\tau+ 2N\cos(\theta)d\phi)^2 + \frac{dr^2}{V(r)} + (r^2-N^2)d\Omega_b^2
\end{equation}
and of
Taub-NUT-AdS and Taub-bolt-AdS spacetimes \cite{PP}, with metrics of the form
\begin{equation}\label{e14}
ds^2= \frac{\ell^2}{4}E\left[\frac{V(r)}{E(r^2-1)}(d\tau+ \sqrt{E}\cos(\theta)d\phi)^2 
+ \frac{4(r^2-1) dr^2}{V(r)} + (r^2-1)d\Omega_b^2\right] \quad .
\end{equation}
In the above, the parameter
$N$ is the NUT charge and $E$ is an arbitrary constant whose significance is discussed below.
The measure
$d\Omega_b^2=d\theta^2+\sin^2(\sqrt{b}\theta)d\phi^2/b$.
For topological black holes \cite{top1,top2} (with compact event horizons of genus $g$),
the parameter $b=1,0,-1$ for $g=0,1,\ge 2$ respectively, the $b=1$ case being the
SAdS metric (\ref{e11}).

Table I summarizes the results, with those for the $\Lambda=0$ spacetimes being obtained
by taking $\ell\to\infty$. The metric function $V(r)$ characterizes the spacetime
under consideration, and the time parameter is periodic, with periodicities as listed.
Provided the counterterm in (\ref{e7}) is included as stated above, all results in table
I are finite.  The NUT solutions in table I have no bolts -- their 
entropy is entirely due to the presence of the Misner string (see ref. \cite{GP} for
an early discussion of the action of Taub-NUT space).

In the Taub-NUT solution the metric has a nut at $r=N$, with a Misner string
running along the z-axis from the NUT out to infinity, and the solution is regular if $r\ge N$.
In the Taub-bolt case the metric is regular provided $r\ge 2N$, with the bolt located at $r=2N$; 
there is also a Misner string
running along the z-axis. The difference between the actions, Hamiltonians and entropies for
these spacetimes yield the results obtained by Hawking and Hunter \cite{Hawkhunt}, 
illustrating that the subtraction procedure they employ yields {\it relative} values for
these quantities. The NUT charge is proportional to the the first Chern
number of the U(1) bundle over the sphere at infinity (in the 
orbit space of the U(1) isometry group): the boundary at infinity is a squashed $S^3$. 

Similarly, in the Taub-NUT-AdS spacetime, $E$ parametrizes the squashing of the 3-sphere at infinity.
In the Taub-bolt-AdS case the squashed $S^3$ has $k$ points identified around the U(1) direction;
the parameter $s$ is arbitary, and $r\ge s$, with the location of the bolt at $r=s$,
and $s>2/k$. For completeness,
the results for Taub-NUT-AdS spacetime with $k$ points identified on the $S^1$ fibre of the squashed
$S^3$ is also given. As with the previous
two cases, it is straight\-forward to show that the differences between the actions, Hamiltonians,
and entropies for the Taub-bolt-AdS and identified Taub-NUT-AdS spacetimes
yields the results of ref. \cite{HPH}.

One of the more unusual results apparent from table I is that in the AdS cases 
the entropy is not positive for all values of the parameters $E$ and $s$. For the Taub-NUT-AdS
case the entropy is positive provided $E<2/3$. The Hamiltonian is positive for $E<1$, so there
exists a range of solutions (with $2/3<E<1$)  for which the Hamiltonian is positive 
but the entropy is negative. For the Taub-bolt 
case, the entropy is positive over the allowed range of $s$ for $k=1,2$, and is positive
for all $k$ provided $s> \sqrt{2+\sqrt{5}}\approx 2.058$. However for $1<s< \sqrt{2+\sqrt{5}}$ 
the entropy is negative, with $\lim_{s\to 1}S = -\infty$ for $k \ge 3$. The Hamiltonian is positive for
$k=1,2$ over the allowed range of $s$ and for $s>\sqrt{3+2\sqrt{3}}\approx 2.542$ for all $k$, but
is otherwise negative, and diverges to $-\infty$ as $s\to 1$. However only for $k = 3$ is there a range
of solutions (with $ 1.595< s < 1.660$ for with positive Hamiltonian and negative entropy;
for all other values of $k$ the entropy is positive provided the Hamiltonian is. The relative
entropy between a bolt spacetime and its periodically matched/identified NUT analogue is always
positive.

The interpretation of these results needs some care. For spacetimes with zero NUT charge,
analytic continuation to the Lorentzian regime is straight\-forward because the boundary at
infinity is the direct product $S^1\times S^2$.  However for non-zero NUT charge this boundary
becomes a squashed $S^3$, and continuation to the Lorentzian regime is problematic.  One
possibility is to interpret the path integral over all metrics as the partition function for
an ensemble of spacetimes with fixed NUT charge.  The prescription (\ref{e7}) removes the
need to make relative comparisons of thermodynamic quantities between spacetimes with similar 
asymptotic properties \cite{Hawkhunt,HPH,Chamb}. However the appearance of negative values
of the Hamiltonian and entropy for certain (rather narrow) parameter ranges in 
the AdS cases is puzzling, and may correspond to some thermodynamic instability. 

Several interesting issues remain.  While it is striking that the single choice 
$A=B=1$ in (\ref{e7}) is sufficient to yield finite
values for all of the quantities computed in table I, the overall consistency and utility
of this prescription (e.g. for non-spherical boundaries) remain open questions. 
From the CFT perspective it is a 
non-analytic function  of the boundary curvature in the $\ell\to \infty$ limit, and an
infinite series in the boundary curvature for $\ell\neq 0$. More generally, the relationship
the results in table I and the partition function and entropy of a conformal field theory
on the boundary remain intriguing issues to explore.

\vskip 0.3 in
The work was supported by a National Science Foundation grant PHY94-07194 and 
by the Natural Sciences and Engineering Research Council of Canada. 
I am grateful to V. Balasubramanian, D. Garfinkle, and S.W. Hawking for interesting discussions,
and to the ITP and to the physics department at UCSB for their hospitality.
After this work was completed I became aware of work by R. Emparan et.al.\cite{EJM}
which overlaps some of the work contained in this paper and arrives at similar conclusions.

\onecolumn
\begin{sideways}
\begin{tabular}{|c|c|c|c|c|c| }\hline
\multicolumn{6}{|c}{\Large Table I}\\
\hline 
& & & & \\
Spacetime & Metric function & Periodicity & Action & $H_\infty$ & Entropy \\ 
 & & & &\\ \hline
 & & & & \\ 
Flat space & $1$ & arbitrary & $0$ & $0$ & $0$ \\ 
             & & & & \\ \hline  
& & & & \\
Schwarzchild & $1-{2m/r}$ & $8\pi m$ & $4\pi m^2$ & $m$ & $4\pi m^2$ \\ 
             & & & & \\ \hline  
& & & & \\
Taub-NUT  & ${(r-N)/(r+N)}$    & $8\pi N$ & $4\pi N^2$ & $N$ & $4\pi N^2$ \\ 
 & & & & \\ \hline
 & & & & \\
Taub-bolt & ${(r-N/2)(r-2N)}/{(r^2-N^2)}$    & $8\pi N$ & $5\pi N^2$ & ${5N}/{4}$ & $5\pi N^2$ \\ 
 & & & & \\ \hline
Topological & & & & \\
 AdS & ${r^2}/{\ell^2}+b-{2M}/{r}$ & $\frac{4\pi\ell^2 r_+}{3r_+^2 + b\ell^2}$ 
          & $v_g \pi r_+^2 \frac{b\ell^2-r_+^2}{3 r_+^2+b\ell^2}$ & $v_g M$ & $v_g \pi r_+^2$ \\
 & & & & \\  \hline 
&$E r^4 + (4-6E)r^2$ & & & \\
Taub-NUT-AdS & $+ 8(E-1)r +4-3E$    & $4\pi\sqrt{E}$ 
     & $\frac{\pi E \ell^2}{2}(2-E)$ & $\frac{\sqrt{E}\ell^2}{2}(1-E)$ 
         & $\pi E \ell^2(1-\frac{3E}{2})$ \\ 
 & & & & & \\ \hline
Identified & ${\cal E} r^4 + (4-6{\cal E})r^2$ & & & \\
Taub-NUT-AdS & $+ 8({\cal E}-1)r +4-3{\cal E}$    & $4\pi\sqrt{{\cal E}}/k$
     &$-\frac{2\pi\ell^2}{9 k}\frac{(ks-2)(3s^2-ks-1)}{(s^2-1)^2} $ & 
  $\frac{\ell^2}{9 \sqrt{{\cal E}} }\frac{(ks-2)(3s^2-2ks+1)}{(s^2-1)^2}$  
         & $\frac{2\pi\ell^2}{3k}\frac{(ks-2)(s^2-ks+1)}{(s^2-1)^2} $ \\ 
 & & & & & \\ \hline
 & ${\cal E} r^4 + (4-6{\cal E})r^2$   & 
     & & &  \\
Taub-bolt-AdS &  $+ [-{\cal E}s^3 +(6{\cal E}-4)s +\frac{3{\cal E}-4}{s}]r$ 
        &$4\pi \sqrt{{\cal E}}/{k}$ &$-\frac{\pi\ell^2}{18 k}\frac{(ks-2)(ks^4-8s^3+3k)}{(s^2-1)^2} $ &
$\frac{\ell^2}{36 \sqrt{{\cal E}} }\frac{(ks-2)(ks^4+4s^3-6ks^2+12s-3k)}{(s^2-1)^2}$ 
& $\frac{\pi\ell^2}{6k}\frac{(ks-2)(ks^4-4ks^2+8s-k)}{(s^2-1)^2} $
\\ 
& $+4-3{\cal E}$ & & & \\
\hline
\multicolumn{3}{|c}{For the Topological }&\multicolumn{3}{c|}{For the} \\
\multicolumn{3}{|c}{SAdS solution:
$M=({r_+^3}/{\ell^2}+br_+)/2$, $\ \ v_g=|g-1|+\delta_{g,1}$}&\multicolumn{3}{c|}{Taub-bolt/NUT-Ads solutions:
${\cal E}=(2ks-4)/(3(s^2+1))$}
\\ 
\multicolumn{6}{|c}{}\\
\hline
\end{tabular}

\end{sideways}

\end{document}